\documentclass[twocolumn,aps,prb,showpacs]{revtex4}
\usepackage{bm}
\usepackage{graphicx}
\usepackage{amssymb}
\usepackage{amsmath}
\usepackage{eufrak}
\usepackage{color}
\newcommand{\nix}[1]{}

\begin{document}


\title{Photon helicity driven electric currents in graphene}

\author{J. Karch$^1$, P. Olbrich$^1$, M.~Schmalzbauer$^1$, C.~Brinsteiner$^1$,
U.~Wurstbauer$^1$, M.\,M.~Glazov$^2$, S.\,A. Tarasenko$^2$, E.\,L. Ivchenko$^2$, D. Weiss$^1$,
J.~Eroms$^1$, and S.\,D.~Ganichev$^1$ }
\affiliation{$^1$  Terahertz Center, University of Regensburg,
93040 Regensburg, Germany}
\affiliation{$^2$ Ioffe Physical-Technical Institute, Russian
Academy of Sciences, 194021 St.~Petersburg, Russia}

\begin{abstract}
We report on the observation of photon helicity driven currents
in graphene. The directed net electric current is generated in single layer graphene by circularly
polarized terahertz laser radiation at normal as well as at oblique incidence and
changes its sign upon reversing the radiation helicity. The phenomenological and
microscopic theories of the observed photocurrents are developed. We demonstrate that
under  oblique incidence the current is caused by the circular photon drag effect  in the  interior of graphene sheet.
By contrast, the effect at normal incidence
stems from the sample edges, which reduce the symmetry and result in an asymmetric
scattering of carriers driven by the radiation field.
Besides a photon helicity dependent current
we also observe photocurrents in response to
linearly polarized radiation. The microscopic mechanisms governing this effect are discussed.
\end{abstract}


\maketitle

\section{Introduction}

Graphene, a one-atom-thick layer of graphite, was experimentally isolated only six
years ago and has since then revealed fascinating effects in a number of
experiments owing to specifics of the electron energy spectrum~\cite{Bib:Novoselov2004}. The chiral motion of charge carriers
leads to a peculiar modification of the quantum Hall effect~\cite{Bib:NovoselovQHE,Bib:KimQHE}
and plays an important role in phase-coherent phenomena such as, e.g., weak localization~\cite{Bib:McCann,Bib:Tikhonenko}.
The fact that the band structure resembles the dispersion relation of a
massless relativistic particle has created enormous excitement since relativistic experiments
in a solid state environment became feasible~\cite{Bib:GeimRise}. Indeed,
Klein tunneling~\cite{Bib:KleinGraphene}, a relativistic effect predicted over
80~years ago~\cite{Bib:Klein}, has been experimentally demonstrated very recently in gated
graphene structures~\cite{Bib:KleinExp}.
Another characteristic feature of graphene is the presence of two valleys, each containing a Dirac cone.
This constitutes a two-state degree of freedom  -- much like the electron spin --  which
was suggested to be applied in valleytronics~\cite{Bib:Valley}.
Most recently several theoretical groups suggested that the combination of 
intense  radiation and a constant electric field applied to single or multilayer graphene may result
in the generation of a valley-polarized~\cite{Yao,Abergela} and anomalous Hall current~\cite{Oka}.

Here we demonstrate that the illumination of
monolayer graphene by radiation of a terahertz (THz) laser in the absence of any $dc$ field applied to the sample
causes directed electric currents, including those solely driven by the radiation helicity.
Photon helicity driven currents are well known in semiconductor low-dimensional
structures, and the photocurrent generation has been proven to be a very efficient method to study
non-equilibrium processes in semiconductors yielding information on their
symmetry, details of the band structure and processes of electron momentum, spin and
energy relaxation, etc.~\cite{review,Ivchenkobook,Ganichevbook,IvchenkoGanichev}.
Microscopic mechanisms of this class of phenomena in quantum wells are based on
spin-orbit coupling in gyrotropic materials~\cite{review,PRL01,Nature02} or on
orbital effects originated from the quantum interference of optical transitions~\cite{Tarasenko07,PRBMOSFET,PRLLATERAL}.
Our experiments evidence that the helicity driven photocurrent in graphene
consists of two contributions. One of them appears at oblique incidence only and is an odd function
of the angle of incidence. The other has its maximum at normal incidence and
is an even function of the incidence angle. We show that the first effect is caused by
the circular photon drag effect~\cite{Ivchenko1980,Belinicher1981,Shalygin2006,gippius09}, see also Ref.~\onlinecite{hertel06}, which stems from the simultaneous transfer
of the linear and angular momenta of photons to the free carriers in the interior of the graphene sheet. The second one,
however, cannot be attributed to any photoelectric effect in an ideal honeycomb lattice of graphene. This two-dimensional lattice possesses a center of space inversion and does not allow for an electric current at normal incidence of the radiation. Thus, the appearance of photocurrents at normal incidence is a clear manifestation of the symmetry reduction of the system.
We suggest that these currents, both the helicity-sensitive and those generated by  linearly polarized radiation, also observed in  experiment,
are caused by the edges of the real finite-size samples.

This paper is organized as follows. In Sec.~\ref{sexperiment}, a short overview of the experimental technique is given.
The experimental results are summarized in Sec.~\ref{sresults}. In Sec.~\ref{stheoryOblique}, we present a phenomenological description and a microscopic theory of the photon drag effect in graphene. In Sec.~\ref{micronormal}, we develop a theory of the edge photogalvanic effect responsible for the observed current at normal incidence of the radiation.


\section{Experiment}
\label{sexperiment}

\begin{figure}[t]
\includegraphics[width=0.95\linewidth]{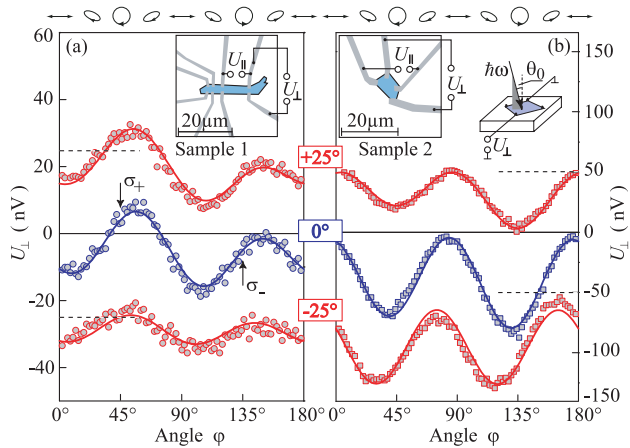}
\caption{Photosignals, $U_\perp$, in a single graphene sheet as a function of the angle $\varphi$, which determines
the radiation helicity.
The data are obtained at room temperature for various angles of incidence, $\theta_0$.
The signal is measured nearly perpendicular
to the light propagation direction applying radiation of the $cw$ THz laser
with the photon energy $10.5$\,meV, power $\approx 20$\,mW and a diameter of the laser spot
about $1$\,mm.
The data for $\theta_0 = \pm 25^\circ$ are shifted by
$\pm 25$~nV [sample $ 1$, panel (a)] and $\pm 50$~nV [sample $ 2$, panel (b)], respectively.
The horizontal dashed lines show $x$-axes corresponding to $U_{\perp} = 0 $ for the shifted data. 
Full lines are fits to Eq.~\protect(\ref{phenom1}). We note that these fits can be obtained
from superposition of the photon drag effect at oblique incidence [see Eqs.~\eqref{j:phi}, \eqref{j:lpd:simp}] and
the photogalvanic effect at normal incidence [see Eq.~\eqref{J_y_3}].
The insets show the sample geometry and experimental configuration.
The ellipses on top illustrate the states of polarization for various angles $\varphi$.
} 
\label{fig1}
\end{figure}

\begin{figure}[t]
\includegraphics[width=\linewidth]{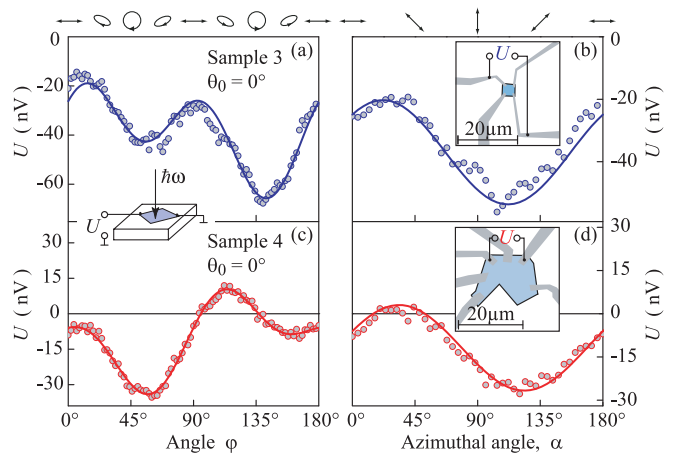}
\caption{Polarization dependences of the photoresponse obtained at normal incidence of the $cw$ THz laser radiation for the samples $3$ and $4$.
The two left panels show signals as a function of the radiation helicity, given by $\varphi$, and the two right panels
as a function of the electric field orientation given by azimuthal angle $\alpha$.
Full lines are fits to
Eq.~\protect(\ref{phenom1}) [panels (a) and (c)] and Eq.~\protect(\ref{phenom2}) [panels (b) and (d)].
These fits can be obtained
by the photogalvanic effect at normal incidence [see Eq.~\eqref{J_y_3}].
The insets show the
experimental arrangement and the samples geometry.
The symbols on top of the left and right panels illustrate the states of polarization for various
angles $\varphi$ and $\protect\alpha$, respectively. 
} \label{fig3}
\end{figure}

The graphene samples were prepared from natural graphite using the mechanical
exfoliation technique~\cite{Bib:Novoselov2004} on an oxidized silicon wafer.
The oxide thickness of $300$\,nm allowed to locate graphene flakes in an optical
microscope and to assess their thickness. We checked the reliability of this method
using Raman spectroscopy and low-temperature quantum Hall measurements on similar samples~\cite{jonathan1}.
Typically samples were $p$-doped by adsorbed contaminants in the range of  
$n \leq 2\times 10^{12}$~cm$^{-2}$.  The Fermi energies were $E_F \leq 165$~meV
and the mobilities at room temperature of the order of $2.5 \times 10^3$~cm$^2$/Vs.
%
The flakes included in this study were all single layer.
After recording the position of the flakes with respect to predefined markers, we contacted
them with electron beam lithography and thermal evaporation of 60 nm Pd electrodes. 
The resistance of graphene measured between various contacts  was in the range of 1~k$\Omega$. 
Some
flakes were cut into shape using oxygen based reactive ion etching.
Insets in Figs.~\ref{fig1} and \ref{fig3} show the shapes and the contact geometry of the four investigated samples
indicated by the numbers $ 1$, $ 2$, $ 3$ and $ 4$.
The samples were glued
onto holders with conductive epoxy enabling the use of the highly doped silicon wafer as a
back gate.
The sample morphology was characterized by atomic force microscopy 
measurements under ambient conditions with the
microscope in intermittent contact mode with standard silicon tips~\cite{jonathan2}.
%
%

The photocurrents were generated at room temperature applying THz radiation
of an optically pumped continuous-wave ($cw$) CH$_{3}$OH laser~\cite{Ganichevbook}
operating at a wavelength $\lambda$= 118\,$\mu$m (or the corresponding photon energy 10.5~meV) with the
power $P \approx $ 20\,mW and a diameter of the laser spot of about 1\,mm.
The radiation was modulated at chopper frequencies in the range from $120$ to $600$\,Hz. 
The sign of the signal is defined as a relative phase with respect to the lock-in reference signal frequency, 
which was kept the same for all measurements.
Additionally we used
a high power pulsed NH$_3$ laser operating at $\lambda = 148\,\mu$m  and $P \approx 30$\,kW\,\cite{Ganichevbook}.
The photocurrent induced by the $cw$ laser is measured across a 10\,M$\Omega$ 
load resistor and recorded with lock-in technique. The signal magnitudes of Figs.~\ref{fig1}-\ref{fig2}
are given before amplification.
The photoresponse to pulsed radiation is measured
across a 50\,$\Omega$ resistor.

The samples were excited both at normal and oblique incidence.
In the case of oblique excitation, the angle $\theta_0$ between the light propagation direction
and the sample normal was varied from $-25^\circ$ to
+25$^\circ$.
The experimental geometries are
illustrated in the insets in Figs.~\ref{fig1} and~\ref{fig3} for oblique and normal incidence, respectively.
In order to vary the radiation helicity we used a
$\lambda$/4 plate. The rotation of the plate resulted in the
change of the degree of circular polarization after $P_{\rm circ}= \sin 2 \varphi$,
where $\varphi$ is the angle between the initial polarization vector of the laser light
$\bm{E}$ and the $c$-axis of the plate.
The light polarization states for certain characteristic angles $\varphi$ are sketched on
top of Figs.~\ref{fig1} and \ref{fig3}.
In some experiments we also used linearly polarized radiation.
In this case the plane of polarization of the radiation
was rotated by
$\lambda/2$ plates. This enabled us to vary the azimuthal angle
$\alpha$ from $0^\circ$ to $180^\circ$ covering all possible
orientations of the electric field vector in the graphene plane.
The relative positions of the light polarization with respect to the
graphene sample edges and contacts are shown on the top of Fig.~\ref{fig3}(b).

\section{Experimental results}\label{sresults}

Irradiating graphene samples at normal as well as at oblique incidence we observed
photocurrents in all investigated samples. The signal was detected at
room temperature for both low power excitation level of $cw$ laser and
high power pulsed laser. In the latter case the width of the
photocurrent pulses was about 100 ns, which corresponds to the THz
laser pulse duration.
In order to prove
that the signal stems from the graphene flakes and not, e.g.,
from the substrate, we removed the graphene layer from one of the
samples and observed that the signal disappeared.

Figure\,\ref{fig1} shows the photoresponse $U_\perp$, which is proportional to the
photocurrent $j$, of two samples\,$1$ and\,$2$ as a function of the angle\,$\varphi$,
assigning the helicity. The signals are measured at
different angles of incidence $\theta_0=0, \pm 25^{\rm o}$.
While the figure shows the photoresponse obtained from the pairs of contacts oriented almost
perpendicular to the light propagation plane, a photocurrent has also been observed at any other pair of contacts.
Generally, the helicity dependence of the photocurrent can be well fitted by
\begin{equation} \label{phenom1}
J = A \sin 2\varphi + B \sin 4\varphi  +  C \cos 4\varphi + D\:,
\end{equation}
%
where $A$, $B$, $C$ and $D$ are fitting parameters.
Such behavior, phenomenologically well described by symmetry arguments
(see below), was found in all graphene samples.
The fits to experimental
data shown by solid lines in Fig.~\ref{fig1} demonstrate a good agreement. The first term on the right hand-side of Eq.~\eqref{phenom1}, which is proportional to $\sin{2\varphi}$ and described by the
parameter $A$, changes its sign upon reversing the photon helicity (marked by arrows). The analysis of the
experimental data gives an evidence
that a substantial contribution to the total photocurrent changes
its sign upon switching the radiation helicity from right- ($\sigma_+$) to left-handed ($\sigma_-$) circularly polarized
light, i.e., for the angles $\varphi = 45^\circ$ and $\varphi = 135^\circ$, respectively. This is most spectacularly seen in the data obtained on sample\,$ 1$,
where the changing sign of the current is directly detected at normal incidence, $\theta_0=0^\circ$ (Fig.~\ref{fig1}a).
The helicity driven photocurrent gives a substantial contribution to the photoresponse
of two other samples as demonstrated in Figs.~\ref{fig3}(a) and (c) for normal incidence of radiation.

\begin{figure}[t]
\includegraphics[width=\linewidth]{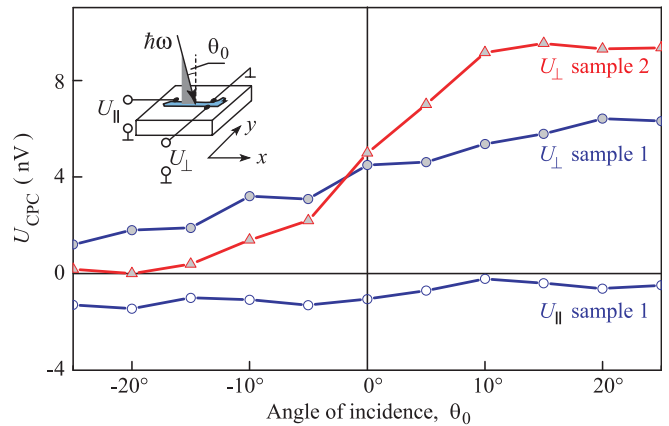}
\caption{
Signals due to circularly polarized radiation $U_{CPC} = [U(\varphi = 45^\circ) - U(\varphi = 135^\circ)]/2$ measured as a function of the incidence angle $\theta_0$
in response to the $cw$ THz laser radiation.
Triangles and full circles show the transversal response, $U_\perp$, for samples $ 1$ and $ 2$, respectively.
Open circles show the signal obtained in the longitudinal geometry, $U_\parallel$, for sample $ 2$.
The inset illustrates the experimental configuration. 
} \label{fig2}
\end{figure}

A deeper insight into the photon-helicity driven photocurrents is given by their
dependences on the incidence angle $\theta_0$ presented in Fig.~\ref{fig2}
for the samples $ 1$ and $ 2$. We obtained that the photoresponse of the helicity driven electric current is substantially different for
the geometries, where the photocurrent is measured in the direction transversal
and almost longitudinal to the light propagation\cite{footnote1}.
While the photocurrent measured in the longitudinal geometry is almost
independent on the angle of incidence and reflects the signals obtained at normal incidence,
in the transverse geometry it shows a superposition of two contributions being even and odd
in the incidence angle. An important feature of the latter contribution
is that its direction is determined by the light propagation plane, the radiation helicity and the sign of $\theta_0$,
irrespective to the orientation of the sample edges by
contrast to the photocurrents being even functions of the angle $\theta_0$.

Now we turn to the photocurrent contributions proportional to the coefficients $B$ and $C$ in Eq.~\eqref{phenom1}.
This photocurrent, in fact, does  not require the radiation helicity and reflects the linear polarization component
of the incident radiation. This has been checked in an independent experiment.
By rotating now a $\lambda/2$ plate we varied the relative position between the plane of
the radiation polarization and the axes of the samples characterized by the azimuthal
angle $\alpha$. The data are shown in Figs.~\ref{fig3}(b) and (d)
and can be well described in agreement with Eq.~\eqref{phenom1} by
\begin{equation} \label{phenom2}
J = 2B \sin 2\alpha +2C\cos{2\alpha} + D - C\:.
\end{equation}
%
Corresponding fits are plotted by the solid lines in Figs.~\ref{fig3}(b) and (d).
Such a behavior of the photocurrent clearly demonstrates the linear-polarization sensitivity of the effects.
Note that the difference $D
- C$ actually constitutes the polarization independent effect, which can be related to e.g. contacts and is outside of the scope of the present paper.
%
%
Studying the dependence of the photocurrent on the angle of incidence   
for linearly polarized excitation we observed that, in contrast
to the excitation with circularly polarized light, the measured
photocurrents both in longitudinal and transverse geometries always result from the superposition of the
contribution weakly dependent on $\theta_0$, which stems from normal incidence, and the
one being odd in the angle $\theta_0$ (not shown).
Moreover, these contributions odd in $\theta_0$ are interconnected: While the longitudinal
one behaves as $\cos{2\alpha}$, the transverse component is proportional to $\sin{2\alpha}$.

To summarize, we observed that both the helicity dependent and helicity independent photocurrents have two contributions:
(i) an even function of the incidence angle $\theta_0$, which dominates the signal at normal incidence,
and (ii) a odd in the incidence angle contribution, which appears at oblique incidence only.
Below in Sec.~\ref{stheoryOblique} we demonstrate that the contributions to the photocurrent odd in the angle $\theta_0$, both circular and linear, stem from
the interior of graphene and are caused by the photon drag effect, where the transfer of the momentum of the photons results in
a directed motion of electrons.
We present the phenomenological approach, which is based on symmetry arguments and is free from
details of the microscopic model, and then the microscopic theory is put forward.
The photoresponse at normal incidence, by contrast, can not be related with an ideal bulk material.
In Sec.~\ref{micronormal}, we show that these effects are due to the sample edges
and give the corresponding microscopic theory.

\section{Theory of photon drag}\label{stheoryOblique}

\subsection{Photocurrents as a second-order response to the electromagnetic fields}
In general, the photocurrents under study can be regarded as a
second-order nonlinear steady-state response
\begin{equation} \label{secondorder}
j_{\lambda}^{(2)}(0,0) = \sigma^{(2)}_{\lambda \nu \eta}(\omega, {\bm q})
E_{\nu}({\bm q}, \omega) E^*_{\eta}({\bm q}, \omega)
\end{equation}
to the electric field of a plane electromagnetic wave
\begin{equation} \label{plane}
{\bm E}({\bm r}, t)= {\bm E}({\bm q}, \omega) {\rm e}^{- {\rm i} \omega t + {\rm i} {\bm q} {\bm r} }
+ {\bm E}^*({\bm q}, \omega) {\rm e}^{{\rm i} \omega t - {\rm i} {\bm q} {\bm r}}\:.
\end{equation}
Here ${\bm E}(\omega, {\bm q})$ is the complex electric-field
amplitude with the frequency $\omega$ and the wave vector ${\bm
q}$, ${\bm j}^{(2)}(0,0)$ is the photocurrent density, zeros
$(0,0)$ indicate that we consider an electric current averaged in
time and space. In a two-dimensional system the index $\lambda$
runs only over the two in-plane coordinates $x$ and $y$ while the
indices $\nu$ and $\eta$ can include the normal coordinate $z$.
The second-order conductivity ${\bm \sigma}^{(2)}(\omega, {\bm
q})$ can be expanded in powers of $\bm q$, hereafter we retain only the
first two terms
\begin{equation} \label{expansion}
\sigma^{(2)}_{\lambda \nu \eta}(\omega, {\bm q}) = \sigma^{(2)}_{\lambda \nu \eta}(\omega, 0)
+ \Phi_{\lambda \mu \nu \eta}(\omega) q_{\mu}\:.
\end{equation}
The tensor $\sigma^{(2)}_{\lambda \nu \eta}(\omega, 0)$ requires a
lack of the inversion center in the symmetry point group of the
system, it can conveniently be decomposed into two tensors ${\bm
\sigma}^{(2,\pm)}$ symmetrical and antisymmetrical with respect to
the index interchange $\nu \leftrightarrow \eta$. The symmetrical
tensor $\sigma^{(2,+)}_{\lambda \nu \eta}$
describes the linear photogalvanic effect (PGE). The
antisymmetric tensor $\sigma^{(2,-)}_{\lambda \nu \eta}$ is dual
to a pseudotensor $\gamma_{\lambda \xi}$ describing the circular
PGE~\cite{Ivchenkobook,Ganichevbook,IvchenkoGanichev,sturman_book}. The contributions described by the
fourth-rank tensor ${\bm \Phi}$ are allowed both in
centrosymmetric and noncentrosymmetric media. In a simplified
qualitative picture the linear-${\bm q}$ contribution in
Eq.~(\ref{expansion}) arises due to transfer of momenta from
photons to free carriers and, therefore, is called the photon drag
effect.

The photon-drag current can equivalently be presented in terms of
spatial derivatives of the electric field ${\bm E}({\bm r},
\omega)$ as follows
\begin{eqnarray} \label{deriv}
&&j_{\lambda}^{({\rm drag})} = \Phi_{\lambda \mu \nu \eta}(\omega)
\\  && \mbox{}\hspace{3 mm} \times \frac{\rm i}{2} \left[ E_{\nu}({\bm r}, \omega) \frac{ \partial
E^*_{\eta}({\bm r}, \omega)}{\partial x_{\mu} } -  E^*_{\eta}({\bm
r}, \omega) \frac{\partial E_{\nu}({\bm r}, \omega)}{\partial
x_{\mu}} \right]\:. \nonumber
\end{eqnarray}
For the plane wave (\ref{plane}) the bilinear expression in the
square brackets is independent of ${\bm r}$. Equation
(\ref{deriv}) permits one to make sure that the tensor
$\Phi_{\lambda \mu \nu \eta}$ in Eq.~(\ref{expansion}) describes
the drag current even in case of the sample dimension $L$ being
smaller than the light wavelength provided $L$ by far exceeds the
carrier free path length $l$.

Similarly to the second-harmonic generation linear in the light
wave vector, there are two mechanisms of the photon drag effect.
The linear-$\bm q$ terms can appear in Eq.~(\ref{expansion}) either
due to the gradient of electric field ($qE^2$-mechanism) or due to
a combined action of the electric and magnetic field of the
electromagnetic wave ($EB$-mechanism or the high-frequency Hall
effect). In some sense, these mechanisms are analogues,
respectively, of the magnetic-dipole and electric-quadrupole
contributions to second-harmonic generation described by the
difference ${\bm j}^{(2)}(2 \omega,2 {\bm q}) - {\bm j}^{(2)}(2
\omega,0)$.

\subsection{Phenomenological description of the drag current in graphene}

The ideal honeycomb lattice of graphene is described by the point
group $D_{6\rm h}$. The group contains the space inversion and,
therefore, allows the photon-drag currents only because the
circular and linear PGE are excluded by symmetry
arguments~\cite{Ivchenkobook}. Decomposing the tensor
$\Phi_{\lambda \mu \nu \eta}$ into symmetrical and antisymmetrical
parts concerning the indices $\nu$ and $\eta$ the drag current can
can be written phenomenologically also as
\begin{equation}
\label{j:pde}
 j_\lambda = T_{\lambda \mu \nu \eta} q_\mu  \frac{e_\nu e_\eta^*+ e_\nu^*e_\eta}{2} I +
 \tilde{T}_{\lambda\mu\nu} q_\mu P_{\rm circ} \hat{e}_\nu I\ \:.
\end{equation}
Here $I$ is the light intensity (energy flux through unit surface)
related with the electric field by $|{\bm E}({\bm q}, \omega)|^2 =
(2 \pi/c n_{\omega})I$, $n_{\omega}$ is the refractive index at
the frequency $\omega$,
\[
T_{\lambda \mu \nu \eta} = \frac{\pi}{c n_{\omega}} (\Phi_{\lambda \mu \nu \eta} +
\Phi_{\lambda \mu \eta \nu})\:,
\]
\[
\tilde{T}_{\lambda \mu \xi} = - \frac{{\rm i} \pi}{2 c n_{\omega}}
\sum_{\nu \eta}{\rm e}_{\xi \nu \eta} (\Phi_{\lambda \mu \nu \eta} -
\Phi_{\lambda \mu \eta \nu}) \:,
\]
${\rm e}_{\xi \nu \eta}$ is the unit antisymmetric tensor,
$\bm{j}$ is the current density, $\bm{e}$ is the unit polarization
vector of the electromagnetic wave, $\hat{\bm e}$ is the unit
vector pointing in the light propagation direction, $P_{\rm circ}$ is
circular polarization degree related to the vectors $\bm{e}$ and
$\hat{\bm e}$ by $P_{\rm circ}\hat{\bm { e}} =  \mathrm i (\bm e
\times \bm e^*)$.

The forth-rank tensor $T_{\lambda \mu \nu \eta}$ describes the
linear photon drag effect which is insensitive to the sign of
circular polarization and reaches its maximum for linearly
polarized light, while the third rank pseudotensor $\tilde
T_{\lambda\mu\nu}$ stands for the circular photon drag current
which changes its sign upon the reversal of photon helicity.

The symmetry analysis shows that, in the $D_{6 \mathrm h}$ point
group, the tensor $T_{\lambda\mu\nu\eta}$ has four linearly
independent components: $T_1 = T_{xxxx} + T_{xxyy}$, $T_2 =
T_{xxxx} - T_{xxyy} = 2 T_{xyxy}$, $T_3 = 2 T_{xzxz}$ and $T_4 =
T_{xxzz}$. Then the first term in the right-hand side of
Eq.~\eqref{j:pde} is reduced to
\begin{subequations}
\label{j:lpde}
\begin{multline}
\label{j:lpde:x}
j_x = T_1  q_x \frac{|e_x|^2 + |e_y|^2}{2} I \\
+ T_2\left(q_x \frac{|e_x|^2 - |e_y|^2}{2}  + q_y \frac{e_xe_y^* + e_x^*e_y}{2}\right) I \\
+ T_3 q_z\frac{e_xe_z^* + e_x^*e_z}{2} I + T_4 q_x |e_z|^2  I \:,
\end{multline}
\begin{multline}
\label{j:lpde:y}
j_y = T_1  q_y \frac{|e_x|^2 + |e_y|^2}{2} I \\
+ T_2\left(q_y \frac{|e_y|^2 - |e_x|^2}{2}  + q_x \frac{e_xe_y^* + e_x^*e_y}{2}\right) I \\
+ T_3 q_z\frac{e_ye_z^* + e_y^*e_z}{2} I + T_4 q_y |e_z|^2  I \:,
\end{multline}
\end{subequations}
where $x$ and $y$ are the axes in the graphene plane, and $z$ is
the structure normal. Equations~\eqref{j:lpde} clearly show that
the photon drag current can be induced only at oblique incidence
of the radiation and vanishes under normal incidence when both the
in-plane component of the photon wave vector $\bm{q}$ and the
out-of-plane component of the polarization vector $\bm{e}$ are
zeros. The photocurrent contains both the longitudinal component
induced in the incidence plane and determined by the parameters
$T_1, T_2, T_4$ and the transversal component proportional to
$T_2$. The current described by the parameter $T_3$ requires
$z$-component of the photon wave vector.

For an elliptically polarized radiation, in particular, for
circular polarization, the photon drag current in structures of
the $D_{6\mathrm h}$ symmetry contains additional contributions
proportional to components of the $\tilde{T}_{\lambda \mu \xi}$
tensor:
\begin{subequations}
\label{j:cpde}
\begin{equation}
\label{j:cpde:x}
j_x = \hspace{3 mm} \tilde{T}_1  q_y P_{\rm circ} \hat{e}_z I - \tilde{T}_2 q_z P_{\rm circ} \hat{e}_y I \:,
\end{equation}
\begin{equation}
\label{j:cpde:y}
\mbox{} \hspace{0 mm}j_y = -\tilde{T}_1 q_x P_{\rm circ} \hat{e}_z I + \tilde{T}_2 q_z P_{\rm circ} \hat{e}_x I \:,
\end{equation}
\end{subequations}
and sensitive to the radiation helicity. Here,
$\tilde{T}_1=\tilde{T}_{xyz}$ and $\tilde{T}_2 = \tilde{T}_{yzx}$
are linearly independent parameters. Similarly to the linear
photon drag effect, the circular photocurrent~\eqref{j:cpde} can
be induced at oblique incidence only. However, in contrast to the
former, the circular photocurrent always flows perpendicularly to
the light incidence plane.

It follows from very general considerations that, in
two-dimensional systems, the contributions from $T_3, T_4$ and
$\tilde{T}_2$ are small and unlikely to be observable. Indeed, in
terms of the semiclassical Boltzmann equation the two-dimensional
carriers are unaffected by the normal component of the electric field $E_z$, provided that the graphene layer is flat, and the
terms in Eqs.~(\ref{j:lpde}), (\ref{j:cpde}) proportional to $e_z,
e^*_z$ or $|e_z|^2$ vanish. For interband quantum optical
transitions, the photocurrent governed by the component $q_z$
related to the currents $\propto q_{x,y}$ has a small parameter
$ka$, where $k$ is the electron wave vector referred to the band
extremum point and $a$ is the well thickness in the semiconductor
quantum wells or the carbon atom radius in the case of graphene.
For intraband indirect optical absorption involving virtual
transtions via other bands the parameter $ka$ is multiplied by
another small parameter $\hbar \omega/\Delta$, where $\Delta $ is
the energy distance to other bands. In graphene $\Delta$ is the
energy difference between the $\pi$ and $\sigma$ orbitals and has an order of several electronvolts.\cite{bassani_book}
Therefore, in the following we will ignore the $T_3, T_4$ and
$\tilde{T}_2$ terms in Eqs.~(\ref{j:lpde}), (\ref{j:cpde}).

Most studies on photon drag effect have been carried out in
crystals with cubic symmetry~\cite{Barlow, grinberg, perel73,
Yaroshetskii80p173, Gibson80p182, ivchenko_ganichev_drag}, simple
metals~\cite{shalaev} and in atomic gases\cite{amusia, dolmatov}.
In these systems the circular photon drag is forbidden because the
symmetry equalizes the coefficients $\tilde{T}_1$ and
$\tilde{T}_2$ in Eq.~(\ref{j:cpde}) so that the photocurrent ${\bm
j}$ becomes proportional to the vector product ${\bm q} \times
\hat{\bm e} = 0$. In  systems of uniaxial symmetry, anisotropic
crystals, quantum-well structures, superlattices etc., the
coefficients $\tilde{T}_1$ and $\tilde{T}_2$ are linearly
independent and allow the circular drag photocurrent. Even more
so, this effect is allowed in graphene where, as mentioned above,
the coefficient $\tilde{T}_2$ is negligible. It is worth
mentioning that the appearance of the transverse linear photon
drag effect in the vicinity of the metal surface was discussed in
Refs.~\onlinecite{goff, gurevich}.

\begin{figure}[t]
\includegraphics[width=0.7\linewidth]{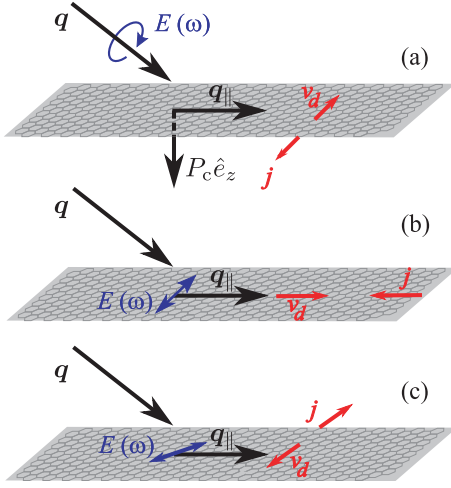}
\caption{Schematic illustration of the photon drag effects.
Panel (a): circular photon drag effect, where the current is generated transverse to the
light incidence plane (see Eqs.~\protect{\eqref{j:cpde}}).
Panels (b) and (c): linear photon drag effect. Here both longitudinal and
transverse current can be generated [see Eqs.~\protect{\eqref{j:lpde}}].
} \label{figt1}
\end{figure}

Figure~\ref{figt1} schematically illustrates the relative
orientation of all remaining contributions in respect to the light
propagation direction and its polarization state. This figure
being in correspondence to Eq.~\eqref{j:cpde} shows that the
helicity driven photocurrent appears at oblique incidence only and
flows perpendicularly to the light propagation direction, see
Fig.~\ref{figt1}(a). Furthermore it changes its sign upon reversal
of the incidence angle. This is in full agreement with the
experiment, see Fig.~\ref{fig2}, where the odd contribution to the
circular photocurrent is indeed observed for the transverse
geometry only. By contrast, the linear photon drag effect is
generated in both longitudinal [see Fig.~\ref{figt1}(b)] and
transversal geometries [see Fig.~\ref{figt1}(c)] being also in
agreement with the experimental findings.

Also, the polarization dependences given by Eqs.~\eqref{j:lpde},
\eqref{j:cpde} are observed experimentally. In order to simplify
the comparison with the experiment we rewrite Eqs.~\eqref{j:lpde} and
\eqref{j:cpde} for the geometry relevant to the experiment with the
light incidence plane being $(xz)$ and the polarization state of
the light described by the angle $\varphi$ for elliptically polarized
light (in particularly, circularly polarized light) and the angle $\alpha$
for linearly polarized light.
In the first case one obtains
\[
j_x(\varphi,\theta_0) \propto \frac{T_1\sin{\theta_0}}{4} [4\cos^2{\theta_0}+\sin^2{\theta_0}(1-\cos{4\varphi})]
\]
\[
 + \frac{T_2\sin{\theta_0}}{4}[4\cos^2{\theta_0}+(1+\cos^2{\theta_0})(\cos{4\varphi}-1)]
\]
\begin{equation}
\label{j:phi1}
j_y(\varphi,\theta_0) \propto \tilde T_1 \sin{\theta_0}\cos{\theta_0}\sin{2\varphi}+\frac{T_2}{2}\sin{\theta_0}\cos{\theta_0} \sin{4\varphi},
\end{equation}
while in the second case one has
\[
j_x(\alpha,\theta_0) \propto T_1\sin{\theta_0}(1-\sin^2\theta_0\cos^2\alpha)
\]
\[
+ T_2\sin{\theta_0}(\cos{2\alpha}-\cos^2{\alpha}\sin^2{\theta_0})
\]
\begin{equation}
\label{j:lpd:simp1}
j_y(\alpha,\theta_0) \propto \sin{\theta_0}\cos{\theta_0}T_2\sin{2\alpha}.
\end{equation}
For the small incidence angles as in our experiment one retains only linear in $\theta_0$ terms and obtains from Eqs.~\eqref{j:phi1}, \eqref{j:lpd:simp1} the following simplified relations describing the photocurrent dependences on $\varphi$:
\[
j_x(\varphi,\theta_0) \propto T_1\sin{\theta_0}
 + \frac{T_2\sin{\theta_0}}{2}(\cos{4\varphi}+1)
\]
\begin{equation}
\label{j:phi}
j_y(\varphi,\theta_0) \propto \tilde T_1 \sin{\theta_0}\sin{2\varphi}+\frac{T_2}{2}\sin{\theta_0} \sin{4\varphi},
\end{equation}
and $\alpha$:
\[
j_x(\alpha,\theta_0) \propto T_1\sin{\theta_0} + T_2\sin{\theta_0}\cos{2\alpha}
\]
\begin{equation}
\label{j:lpd:simp}
j_y(\alpha,\theta_0) \propto \sin{\theta_0}\cos{\theta_0}T_2\sin{2\alpha}.
\end{equation}
The dependence of the experimentally observed 
photocurrents odd in the angle incidence on the parameters $\alpha$ and $\varphi$
characterizing the polarization states of the radiation is in
agreement with Eq.~\eqref{j:phi} and \eqref{j:lpd:simp}. Indeed, as
it is seen from Fig.~\ref{fig1} (also from Fig.~\ref{fig2}) the odd
in $\theta_0$ contribution to helicity driven photocurrent,
$\propto \sin{2\varphi}$, is observed only in the transversal with
respect to the light incidence plane geometry. By contrast, the contribution to the current
odd in the incidence angle, which was measured in
the longitudinal geometry, is described by $\cos{4\varphi}$ or
$\cos{2\alpha}$ (not shown) demonstrating the linear photon drag
effect.

%
%

It is worth mentioning that the coefficients describing the linear
and circular drag effects have different properties under time
reversal, $t\to -t$. Namely, the circular polarization sign
changes at time reversal while the linear polarization sign remains.
Both the electric current and the photon wave vector are odd
functions with respect to the replacement $t\to -t$, therefore the
constants $\tilde{T}_1$ and $\tilde{T}_2$ are proportional to an
odd number of dissipative parameters and the constants $T_i$
($i=1,\ldots, 4$) to an even number of dissipative parameters.

The above analysis, which yields a good agreement with the
experiment, was performed using the ideal symmetry for the
graphene layer characterized by the point group $D_{6\rm h}$. The
real structures, however, are deposited on a substrate, which
removes the equivalence of the $z$ and $-z$ directions and reduces
the symmetry to the $C_{6\rm v}$ point group. Our analysis
demonstrates that the photon drag effect does not change
qualitatively with the reduction of symmetry and only the values
of the parameters $T_1, \ldots, T_4$, $\tilde T_1$, and $\tilde
T_2$ may change. However, now the system lacks an inversion
center, and photogalvanic effects become possible. The
photogalvanic contributions to the total electric current are
phenomenologically given by
\begin{equation}
\label{j:pge}
j_\lambda = \chi_{\lambda\mu\nu} \frac{e_\mu e_\nu^*+ e_\mu^*e_\nu}{2}I +
\gamma_{\lambda\mu} P_{circ} \hat{ e}_\mu I \:.
\end{equation}
In structures of the $C_{6\rm v}$ point group, non-zero components
of the tensors $\chi_{\lambda\mu\nu}$ and $\gamma_{\lambda\mu}$
describing the linear and circular photogalvanic effects,
respectively, are $\chi_{xxz} = \chi_{yyz}$ and
$\gamma_{xy} = -\gamma_{yx}$. It means that the linear
photogalvanic effect requires a $z$-component of the electric field.
Moreover, the circular photogalvanic effect also needs this
component because, e.g., $\hat{e}_x \sim \mathrm i (e_x
e_z^*-e_ze_x^*)$ vanishes if the $z$ component of the field is
absent.
These requirements result in vanishingly small contribution of the
photogalvanic effects to the total electric current, similar to
the arguments, which exclude photon drag effect components
given by the constants $T_3$, $T_4$ and $\tilde T_2$.

To summarize, the analysis of the photocurrents odd-in $\theta_0$
can be reduced to the photon drag effects solely, and can be
applied to the experimental data independent of the influence of
the substrate indicated above for the symmetry of the graphene
flakes.

\subsection{Microscopic theory
}\label{microoblique}

Now we turn to the microscopic theory of both linear and circular
photon drag effect. In the following we focus on the classical
regime of interaction between the radiation and the electron
ensemble in graphene which is realized as soon as $ \hbar \omega
\ll \bar{\cal E}$, where $\bar{\mathcal E}$ is the characteristic
electron energy, the Fermi energy in the degenerate electron gas
or thermal energy $k_B T$ for non-degenerate electrons ($k_B $ is
the Boltzmann constant). In this case only intraband transitions
are involved. This approach is similar to the classical
consideration of the drag effect in
semiconductors~\cite{Barlow,grinberg,perel73,Yaroshetskii80p173,Gibson80p182}.
The theory for the drag effect in the quantum range of
frequencies, $\hbar \omega \sim \bar{\mathcal E}$ or higher,
involving both interband and intraband optical transitions is out
of scope of this paper and will be presented
elsewhere~[\onlinecite{git_drag_unp}]. 

\subsubsection{Qualitative picture} 
\begin{figure}[t]
\includegraphics[width=0.8\linewidth]{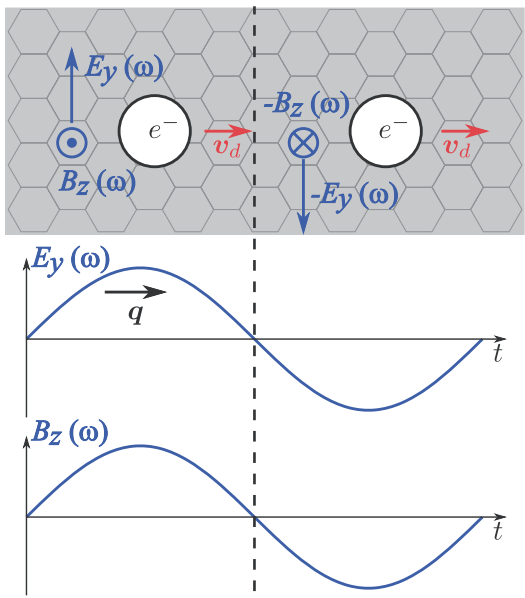}
\caption{Schematic illustration of the high-frequency Hall effect resulting in the photon drag effect.
}
\label{fig:theor:hfh}
\end{figure}

First we present a qualitative microscopic picture of the
drag effect exclusively based on Newton's second law of motion
\begin{equation} \label{Newton}
\frac{d \bm p}{d t} + \frac{\bm p}{\tau} = e
{\bm E}({\bm r}, t) +  \frac{e}{c} [{\bm v} \times {\bm B}({\bm r}, t)]
\:,
\end{equation}
where $e = - |e|$ is the electron charge, $\bm p$ and $\bm v$ are
the electron momentum and velocity, ${\bm p}/\tau$ is the friction
force due to electron scattering with $\tau$ being the scattering
time.  Assuming for simplicity the propagation of
linearly-polarized radiation along the graphene (grazing
incidence) with in-plane wave vector and electric field and
out-of-plane magnetic field: ${\bm q}
\parallel x$, ${\bm E}
\parallel y$, ${\bm B}
\parallel z$, as shown in Fig.~\ref{fig:theor:hfh}. The generation
of a $dc$ current can be described in the framework of the
high-frequency Hall effect~\cite{Barlow}. It is seen from
Fig.~\ref{fig:theor:hfh}(a) that during the first half of the
period $T_{\omega}= 2 \pi/\omega$ of the electro-magnetic field
oscillation where the both field components $E_y$ and $B_z$ are
positive the electron is exposed to action of the two forces $F_y
= e E_y$ and $F_x = e v_y B_z$  like in the Hall effect. In the
absence of friction the phase shift between oscillations of the
velocity $v_y$ and the field $B_z$ equals $90^{\circ}$ and the
average value of $F_x$ vanishes. Allowance for the friction
results in a drift along the $x$ axis. In the second half-period
$T_{\omega}/2$ both fields $E_y$ and $B_z$ simultaneously
reverse and the drift velocity retains its direction. As a result,
a non-vanishing time-averaged drag current is induced in the
direction of $\bm q$. Obviously, the reversal $\bm q \to - \bm q$
results in a change of relative sign of the electric- and
magnetic-field components and, consequently, in the current
reversal. This is the well-known longitudinal linear photon drag
effect. In graphene, as distinct from bulk cubic crystal, the
photon-drag mechanism due to the high-frequency Hall effect has a
characteristic polarization dependence, see Eqs.~(\ref{j:lpde})
for $T_3=T_4 =0$. Indeed, if the electric field of the
grazing-incidence radiation is rotated around ${\bm q}$ by an
angle $\alpha$, then its in-plane component varies $\propto
\cos{\alpha}$, the out-of-plane component of the magnetic field
$B_z$ behaves just as $E_y$, and the drag current $j_x \propto
e^2_y$ decreases by $\cos^2{\alpha} = (1 + \cos{2\alpha})/2$. The
geometry of incidence in the plane $(x,z)$ at arbitrary oblique
angle $\theta_0$ is analyzed in the same way. For the
$s$-polarized light, the consideration is completely the same, a
decrease of $B_z$ by a factor of $\sin{\theta_0}$ is automatically
reproduced by the corresponding change of $q_x = |{\bm q}|
\sin{\theta_0}$. For the linear polarization with the angle
$\alpha$ different from integer multiples of 90$^\circ$, the both
components $E_x$ and $E_y$ are non-zero which yields a
$y$-component of the drag current proportional to $\sin{\theta_0}
\sin{2\alpha}$.

For the $EB$-mechanism, we could ignore the spatial dependence of
the electric and magnetic fields. While considering the
$qE^2$-mechanism we can ignore the magnetic field in the
right-hand side of Eq.~(\ref{Newton}) but take into account the
first-order spatial correction and present the electric field of
the linearly polarized light as
\begin{equation} \label{qxx}
{\bm E}({\bm r},
\omega) = 2 {\bm E}_0 [ \cos{(\omega t)} - 2 q_x x
\sin{(\omega t)}]\:,
\end{equation}
where ${\bm E}_0$ is the electric field amplitude assumed, without
loss of generality, to be real. At the first step we find the
linear response
\begin{equation} \label{pt}
{\bm p}(t) = e \tau {\bm
E}_{0\parallel}{\rm Re}  \left( \frac{{\rm e}^{- {\rm i} \omega t}}{1 - {\rm i} \omega \tau }
\right)
\end{equation}
and the corresponding velocity ${\bm v}(t)$, where ${\bm E}_{0
\parallel} $ is the in-plane component of the vector ${\bm
E}_0$. At the next step we calculate the oscillation of the
electron position $x(t) = \int v_x(t)dt$. Finally, the function
$x(t)$ is substituted into the second term of Eq.~(\ref{qxx}) and
and then the time-average of this term is taken. As a result we
obtain an additional contribution $\propto q_x E_0^2$ to the
continuous force acting on the electron.

The circular photon drag
effect~\cite{Ivchenko1980,Belinicher1981,Shalygin2006} can be
described in a similar way. However, in this case one should take
into account in Eq.~(\ref{Newton}) not only the Lorentz force or
the spatial gradient of the electric field but also the phase
shift by $90$ degrees of the two orthogonal oscillating
electric-field components ${\bm E}_1 \perp {\bm E}_2 \perp {\bm
q}$. Then one can show that, since the phase of the oscillation
${\bm p}(t)$ is retarded in time by $\arctan(\omega \tau)$
relative to ${\bm E}(t)$, see Eq.~(\ref{pt}), the circularly
polarized electromagnetic wave as well induces the drag current
which is perpendicular to ${\bm q}_{\parallel}$. The sign reversal
of radiation helicity means a change $90^{\circ} \to - 90^{\circ}$
of the phase shift between the ${\bm E}_1$ and ${\bm E}_2$
components and a reversal of the circular drag current.

The performed qualitative consideration clearly confirms the
statement made in Sect.~IV.A that the interpretation of photon
drag current in study is independent of the relation between the
sample linear in-plane dimension, $L$, and the light wave length
$\lambda$ but requires but requires a small ratio $l/L$, where $l$
is the mean free path length $v\tau$, as well as a small product
$q l$.

\subsubsection{Boltzmann equation treatment}

We develop a microscopic kinetic theory of the photon drag effect
for the simplest model of the graphene band structure. We consider
only the conduction and valence band states in the $\bm K$ and
$\bm K'$ valleys formed from the $\pi$-orbitals of carbon atoms.
In each valley, the effective Hamiltonian is a 2$\times$2 matrix
\begin{equation}
\label{ham}
\mathcal H = \hbar v (\bm k \cdot \bm \sigma)\:,
\end{equation}
which describes both the conduction and valence bands of a
graphene layer. Here $v$ is the electron speed in graphene, $\bm k
= (k_x,k_y)$ is the two-dimensional wave vector referred to the
Dirac point, $\bm \sigma$ is a matrix two-dimensional vector with
components $\sigma_x$ and $\sigma_y$ being the Pauli matrices in
the pseudospin space.

In accordance with the definition (\ref{plane}) the electric and
magnetic fields of the incident wave are taken in the form
\begin{equation}  \label{emf}
\bm E = \bm E_0 \mathrm e^{\mathrm i (\bm q\bm r - \omega t)} + {\rm c.c.}\:, \quad \bm B
= \bm B_0 \mathrm e^{\mathrm i (\bm q\bm r - \omega t)} + {\rm c.c.}\:,
\end{equation}
where $\bm E_0$ and $\bm B_0$ are the complex amplitudes which are
orthogonal to each other and to the wave vector $\bm q$. In the
case of a linearly polarized light the vectors ${\bm q}, {\bm E}$
and ${\bm B}$ form a right-hand triple.
It is worth noting that the electric and magnetic fields acting on
an electron in graphene lying on a substrate are different from
those of an incident wave owing to the presence of substrate. To
be specific, let us assume that $(xz)$ is the radiation incidence
plane. Therefore, $E_{0,y} = E_{0,y}^{(i)}(1 + r_s)$ and $E_{0,x}
= E_{0,x}^{(i)}(1 + r_p)$, where the superscript $(i)$ denotes the
incident waves and $r_s$, $r_p$ are the amplitude reflection
coefficients in the $s$ and $p$ polarizations, respectively.
Similar relation holds for the $z$-component of the magnetic
field, $B_{0,z} = B_{0,z}^{(i)}(1+r_s)$. We expect that the
metallic contacts attached to our small samples cause an
inessential distortion of the plane-wave character of the
electromagnetic field.

Following Ref.~\onlinecite{perel73} the kinetic equation for the
electron distribution function $f(\bm k, \bm r, t)$ in a given
valley reads
\begin{equation}
 \label{kinetic:gen}
\frac{\partial f}{\partial t} + \bm v \frac{\partial f}{\partial \bm r} +
\frac{e}{\hbar}\left(\bm E + \frac{1}{c}[\bm v \times \bm B]\right)
\frac{\partial f}{\partial \bm k} = Q\{f\}\:.
\end{equation}
Here $c$ is the light speed in vacuum, $\bm v = \bm v_{\bm k}$ is
the electron velocity in the state with the wave vector $\bm k$,
\begin{equation}
 \label{vel:def}
\bm v_{\bm k} = \frac{1}{\hbar} \frac{\partial \varepsilon_k}{\partial \bm k}
= v\frac{\bm k}{|\bm k|}\:,
\end{equation}
$\varepsilon_k = v \hbar k$ is the electron dispersion, and
$Q\{f\}$ is the collision integral.

The drag current is proportional to the light intensity and,
therefore, appears in the second order in the electromagnetic
fields. Correspondingly, we solve Eq.~\eqref{kinetic:gen} by
iterations with respect to the electric and magnetic fields and
represent the electron distribution function as
\begin{eqnarray}
f(\bm k, \bm r, t) = f_0(\varepsilon_k) \hspace{4 cm} \\
+ \left[ f_1(\bm k)
\mathrm e^{\mathrm i (\bm q \bm r - \omega t)} + {\rm c.c.} \right]+ f_2 (\bm k)+  \ldots, \nonumber
\end{eqnarray}
where $f_0(\varepsilon_k)$ is the equilibrium distribution
function, $f_1(\bm k)$ describes the linear response to the
fields, $f_2(\bm k)$ is a homogeneous time-independent correction
which appears in the second order in $\bm E, \bm B$, and the
omitted terms ($\ldots$) describe other contributions including
those oscillating with a double frequency and terms of the higher
order in $\bm E,\bm B$. The direct current caused by the photon
drag effect is given by
\begin{equation}
\label{jdc}
\bm j = 4e\sum_{\bm k} \bm v_{\bm k} f_2(\bm k)\:,
\end{equation}
where the factor $4$ accounts for the spin and valley degeneracy.

The first-order correction $f_1(\bm k)$ can be found by solving
the linearized Eq.~\eqref{kinetic:gen} with the result
\begin{multline}
\label{f1k}
f_1(\bm k) = - \frac{e\tau_1E_0f_0' }{1- \mathrm i \omega \tau_1}\\
 \times\left[(\bm e \bm v)  - \mathrm i \tau_2\frac{ (\bm q \bm v)(\bm e
\bm v) -v^2(\bm q\bm e)/2}{1-\mathrm i \omega\tau_2} +\frac{(\bm
q\bm e)v^2}{2\omega}\right]\:.
\end{multline}
Here, $f_0'= df_0/d\varepsilon_k$, $\tau_1$ is the momentum
relaxation time (more precisely, relaxation time of the first
angular harmonic of the distribution function) and $\tau_2$ is the
relaxation time of the second angular harmonic. While deriving
Eq.~\eqref{f1k}, we took into account that $\tau_2 |\bm q| v, |\bm
q| v/\omega \ll 1$, in which case the second and third terms in
the square brackets are small corrections to the first one.
Moreover, the energy relaxation time $\tau_{\varepsilon}$
(relaxation time for the zeroth angular harmonics of the
nonequilibrium distribution function) was assumed to be much
longer than all other time scales in the system, namely, $\tau_1,
\tau_2, 1/\omega$. The latter assumption allows us to neglect
corrections to the distribution function related to energy
relaxation processes and allow for this relaxation time to be
infinite~\cite{perel73}. In this connection we note that the first
term in Eq.~\eqref{f1k} contains the first-order angular harmonics
of the wave vector, the second term contains the
second-order harmonics while the third term is angle independent.

The second-order correction $f_2(\bm k)$ averaged in time and
space is found from the equation
\begin{equation}
 \label{f2k}
\frac{2e}{\hbar} {\rm Re}{\left\{\left(\bm E_0 + \frac{1}{c}
[\bm v \times \bm B_0] \right) \frac{\partial f_1^*}{\partial \bm k}\right\}} = Q\{ f_2\},
\end{equation}
where the asterisk denotes the complex conjugation. There are two
contributions to $f_2(\bm k)$ related to the above-mentioned $EB$-
and $qE^2$-mechanisms: the first arises from a product of the
Lorentz force and the main term in Eq.~\eqref{f1k}, and the second
results from a product of the electric-field force and the
$\bm q$-dependent terms in Eq.~\eqref{f1k}.


\begin{figure}[t]
\includegraphics[width=0.8\linewidth]{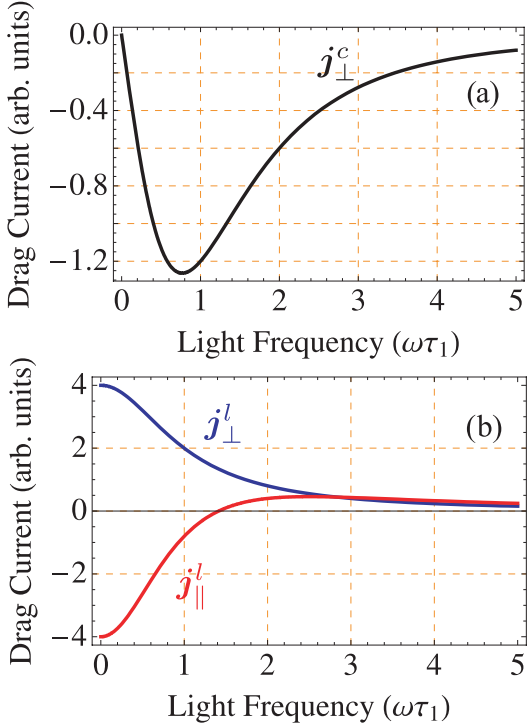}
\caption{Theoretical frequency dependences of (a) the circular photon-drag current,
$j_\perp^c$, and (b) the linear photon-drag currents, both
longitudinal ($j_\parallel^l$) and transversal ($j_\perp^l$),
induced in bulk graphene sheet at oblique incidence and calculated after Eq.~\eqref{classical},
\eqref{classical:cde}. The drag current is given in arbitrary units and
the frequency is presented in the dimensionless form as $\omega\tau_1$, where $\tau_1$ is the momentum
relaxation time.
} \label{figt2}
\end{figure}

Substituting the solution of Eq.~\eqref{f2k} into Eq.~\eqref{jdc}
we arrive at the following expressions
\begin{subequations}
\label{classical}
\begin{multline}
 T_1 = -\frac{4\pi e^3v^4}{\omega c} \sum_{\bm k} \frac{\tau_1  f_0'}{1+\omega^2\tau_1^2} \times
 \\
  \left[2\left( \frac{d \tau_1}{d
 \varepsilon_k} + \frac{\tau_1}{\varepsilon_k} \right) -\frac{1-\omega^2\tau_1\tau_2}{1
 +\omega^2\tau_2^2}\left(\frac{d \tau_1}{d\varepsilon_k} - \frac{\tau_1}{\varepsilon_k}  \right)\right],
 \end{multline}
 \begin{equation}
 T_2 = -\frac{4\pi e^3v^4}{\omega c} \sum_{\bm k} \frac{\tau_1  f_0'}{1+\omega^2\tau_1^2}
 \left(\frac{d \tau_1}{d\varepsilon_k} - \frac{\tau_1}{\varepsilon_k} \right)
\end{equation}
\end{subequations}
for the constants $T_1$ and $T_2$ describing the linear
photon-drag current in the phenomenological
equations~(\ref{j:lpde}).
It is worth noting
that in the low frequency limit, $\omega\tau_1,\omega\tau_2 \ll
1$, the coefficients $T_1$, $T_2$ tend to infinity as
$\omega^{-1}$. However the drag current has a finite limit at
$\omega \to 0$ since it is proportional to $q T_{1,2}$ and $q
\propto \omega$.
It
should also be pointed out that, according to the definition
(\ref{emf}), the static fields $E$ and $B$ differ by factors of 2
from the amplitudes $E_0$ and $B_0$ of the corresponding fields
taken at $\omega \to 0$ and, therefore, the constant relating the
current at $\omega \to 0$ with the squared amplitude $E^2_0$ and
the constant relating the static Hall current with the product of
static fields $E$ and $B$ differ by a factor of 4.

In the same way we derive the constant $\tilde T_1$ describing in
Eqs.~(\ref{j:cpde}) the circular photon-drag effect
\begin{equation}
\label{classical:cde}
\tilde{T}_1 = \frac{2\pi e^3 v^4}{c}\sum_{\bm k}
\frac{\tau_1^2(1+\tau_2/\tau_1)f_0'}{[1+(\omega \tau_1)^2][1 + (\omega \tau_2)^2]}
\left( \frac{d \tau_1}{d \varepsilon_k} - \frac{\tau_1}{\varepsilon_k} \right)\:.
\end{equation}
It is finite at $\omega \to 0$ which means the vanishing of the
circular photocurrent in the static-field limit as expected
because, for static fields, ellipticity is forbidden.

The time inversion symmetry imposes the certain restrictions on
the coefficients in Eqs.~(\ref{j:lpde}) and (\ref{j:cpde}):
microscopic expressions for $T_j$ ($j=1...4$) and $\tilde{T}_j$
($j = 1,2$) must be proportional, respectively, to even and odd
number of dissipative constants, in our case the relaxation times
$\tau_1$, $\tau_2$ or the inverse times $\tau^{-1}_1$,
$\tau^{-1}_2$. One can see that Eqs.~(\ref{classical}) and
(\ref{classical:cde}) for the linear and circular drag effects
satisfy these general rules.

Thus, we have derived microscopic expressions for the
phenomenological constants in Eqs.~\eqref{j:lpde:x},
\eqref{j:lpde:y}, Eqs.~\eqref{j:cpde:x} and \eqref{j:cpde:y}
which, as we addressed above, describe well all experimental
findings. The developed microscopic theory yields all allowed
contributions to the photon drag effect in the model where only
$\pi$-orbitals of carbon atoms are taken into account.

Figure~\ref{figt2} shows calculated frequency dependences of the
photocurrent. In the calculation we assume that the electron
scattering in graphene is short-range, in which case one has
$\tau_1 = 2 \tau_2 \propto \varepsilon_k^{-1}$. It is seen that at
small frequencies, $\omega\to 0$, the circular photon drag effect
vanishes and the linear photocurrent reaches its maximum values.
The absolute value of $j^c$ has a maximum at $\omega \tau_1
\approx 0.8$ and decreases as $\omega^{-4}$ at high frequencies,
$\omega \tau_1, \omega \tau_2 \gg 1$. The linear photon-drag
current exhibits a decrease at high-frequency $\propto \omega^{-3}$.

Finally, we note that together with electrons in the conduction
band, holes in the valence band also contribute to the
photocurrent. The hole contribution to the current is obtained
from Eqs.~(\ref{classical}), \eqref{classical:cde} by the
replacement $e \rightarrow -e$ and the equilibrium electron
distribution function $f_0$ by the hole distribution $f_0^{({\rm
hole})}$.
\section{Edge photocurrents}\label{micronormal}
Here we develop a theory of the photocurrent generation in
graphene at normal incidence. In bulk graphene sheets of the
$D_{6\rm h}$ or $C_{6\rm v}$ symmetry arguments the effect is
forbidden. In order to explain its appearance in experiment, we
need to take into account the graphene edges which locally reduce
the symmetry\cite{footnote2}.
\subsection{Model}

\begin{figure}[t]
\includegraphics[width=0.9\linewidth]{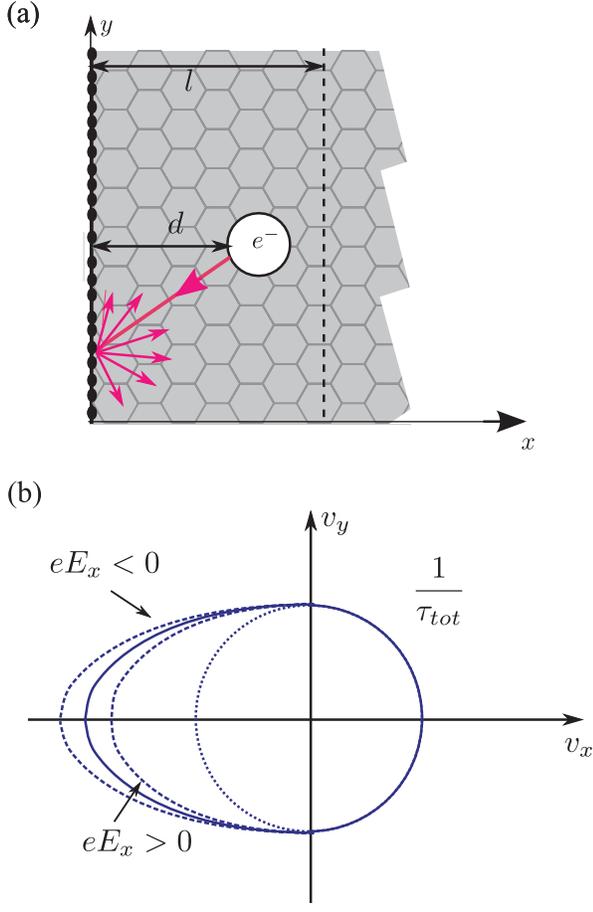}
\caption{Schematic illustration of the edge current generation in a sample of finite size.
}
\label{fig:theor:edge}
\end{figure}
The existence of edge currents induced both by linearly or
circularly polarized radiation can be understood in the framework
of the model depicted in Fig.~\ref{fig:theor:edge}. We will show
that the current can be generated along the sample edge in a
narrow channel of the width comparable to the mean free path $l$.
Then, if the opposite sample edges, in the following the left and
right edges, are non-equivalent, the edge currents do not
compensate each other resulting in a non-zero net electric
current. Otherwise, currents flowing along the opposite edges are oppositely directed and have the same absolute values. In the simplest model of the edge inequivalence, the
electrons are specularly reflected from the right edge of the
sample and diffusively scattered by the left one, as illustrated
in Fig.~\ref{fig:theor:edge}. In a more general model, we can
separate the edge reflectivity into two parts, specular $R_{\rm
sp}$ and diffuse $R_{\rm dif}$, satisfying the identity $R_{\rm
sp}+R_{\rm dif} = 1$. In this model the net current is
proportional to the difference $R_{{\rm dif}, L} - R_{{\rm dif},
R}$ of the left- and right-edge diffusivenesses. For distinctness,
we consider the current generation near the sample left edge and
refer to Fig.~\ref{fig:theor:edge}. An important point to bear in
mind is that an electron traveling in the $l$-thick stripe
adjacent to the edge randomly changes its direction of motion
along the $y$ axis not only in the bulk scattering processes
described by the relaxation time $\tau_1$ but also in the diffusive
reflection at the edge. Thus, its total scattering rate is a sum
of two terms,
\begin{equation}
\label{tau:tot}
\frac{1}{\tau_{\rm tot}} = \frac{1}{\tau} + R_{{\rm dif}, L} \frac{|v_x|}{d} \Theta(-v_x)\:,
\end{equation}
where $d$ is the distance from the electron to the left edge, see
Fig. ~\ref{fig:theor:edge}, and $\Theta(v)$ is the Heaviside
function. The factor $\Theta(-v_x)$ ensures that the additional
contribution to $\tau^{-1}_{\rm tot}$ exists only for the carrier
moving towards the edge. The dependence of the scattering time
$\tau_{\rm tot}$ on the velocity direction is plotted in
Fig.~\ref{fig:theor:edge}(b).

The $x$-component of electric field results in the oscillating
correction to the electron velocity $x$-component:
\begin{equation}
\label{velocity}
v_x = v_x^0 + \delta v_x,
\end{equation}
where
\begin{equation}
\label{dvx}
\delta v_x \sim \frac{e\tau E_x v}{(1-\mathrm i \omega \tau)p} \mathrm e^{-\mathrm i \omega t}
+ {\rm c.c.},
\end{equation}
where $p$ is the electron momentum corresponding to its mean
kinetic energy. The modulation of the velocity results in the
modulation of the scattering time $\tau$ in agreement with
Eq.~\eqref{tau:tot}, since electron reaches the boundary faster of
slower depending on the value of $eE_x$, see
Fig.~\ref{fig:theor:edge}(b).

The $dc$ current along the edge is generated by $y$ component of
the field and, according to Drude theory, is proportional to the
time-averaged product of $E_y\tau_{\rm tot}$:
\begin{equation}
\label{Jy:tot}
J_y \sim e^2 N l \frac{v}{p} {\rm Re}\{\overline{ E_y^*\delta \tau\}}.
\end{equation}
Here $N$ is the electron concentration, $Nl$ is the electron
density per unit length in the vicinity of the edge, overline
means the averaging over time,
\begin{equation}
\label{delta:tau}
\delta\tau = \tau_{\rm tot} - \tau \sim \Theta(v_x) \frac{\delta v_x}{l} \tau^2.
\end{equation}
is the oscillating correction to the scattering time, and, for
simplicity, we assumed that $R_{{\rm diff}, L} \sim 1$. The value
and direction of the $dc$ current is controlled by the relative
phase of the $E_y$ and $\tau_{\rm tot}$, or accordingly $v_x$
oscillations. At $\omega\tau\ll 1$ the oscillations of $E_x$ and
$v_x$ are in phase (no retardation) and the $dc$ current appears
at linear polarization of incident radiation where $E_x$ and $E_y$
components oscillate in phase. With an increase of $\omega\tau$
the retardation effect comes into play and, in addition to the
linear photocurrent, a helicity dependent response appears: the
oscillations of $\tau_{\rm tot}$ appear to be phase-shifted with
respect to the oscillations of $E_x$ and for the circularly
polarized wave where $E_x$ and $E_y$ are $\pi/2$ phase-shifted the
$dc$ current is generated. Taking the average one can find that
\begin{equation}
\label{LPGE}
\mbox{LPGE:} \quad J_y^{\rm lin} \sim \frac{e^3v^2}{p^2} (E_xE_y^*+E_yE_x^*) \tau^3 N ,
\end{equation}
\begin{equation}
\label{CPGE}
\mbox{CPGE:} \quad J_y^{\rm circ} \sim \omega\tau \frac{e^3v^2}{p^2} \mathrm i(E_xE_y^*-E_yE_x^*) \tau^3 N,
\end{equation}
at $\omega\tau\ll 1$.\cite{footnote3} Superposition of these two
current contribution yields polarization dependence of the
photocurrent observed in experiment. Indeed, in the experimental
geometry, where elliptically polarized radiation is obtained by
rotation of the $\lambda/4$ plate, polarization determined terms
reduces to $E_xE_y^*+E_yE_x^* = B \sin(4 \varphi) + C \cos(4
\varphi)$ and $\mathrm i(E_xE_y^*-E_yE_x^*) = A \sin 2\varphi$ being in
agreement with experimental findings (see Fig.~\ref{fig1} and
\ref{fig3}). Also in the set-up, where linear polarization is
rotated by the angle $\alpha$, we obtain from Eq.~(\ref{LPGE})
$E_xE_y^*+E_yE_x^* = 2B \sin(2 \alpha) + 2C \cos(2 \alpha)$
which also agrees with experiment (see Fig.~\ref{fig3}).

As we mentioned above, different edges of the sample make
different contributions to the photocurrents. Consider the
square-shaped sample and assume that all the sample edges scatter
electrons diffusively. Obviously, the currents flowing along the
opposite edges of the sample should have opposite directions. The
current topology depends on the radiation polarization state. In
accordance with our model, the helicity driven current  should
wind in the same direction, clock-wise or counter-clock-wise
depending on the radiation helicity, for all the sample edges
since the helicity, $\mathrm i (E_xE_y^*-E_yE_x^*) $ is preserved
at the rotation by $90^{\rm o}$. Hence, the distribution of
helicity driven current should form a vertex, whose winding
direction changes with the change of light helicity.
However, it is not the case for the
photocurrent caused by linearly polarized light and described
by the combination $E_xE_y^*+E_yE_x^*$. The latter changes its
sign at the $90^{\rm o}$ rotation, hence, the linear photocurrents
flow towards (or outwards) the same corner for adjacent edges.

\subsection{Microscopic theory}

The microscopic theory of this effect is developed in the
framework of kinetic equations for the classical frequency range.
The equation for the distribution function $f(\bm{k},x,t)$ of
electrons in the conduction band in a semi-infinite sample ($x
\ge 0$) has the form
\begin{equation}\label{f_general}
\frac{\partial f}{\partial t} + v_x  \frac{\partial f}{\partial x}      + \frac{e \bm{E}(t)}{\hbar} \frac{\partial f}{\partial \bm{k}} = Q\{f\} \:,
\end{equation}
where $\bm{E}(t) = \bm{E}_0 e^{- \mathrm i \omega t} + \bm{E}_0^* e^{+\mathrm  i \omega t}$ is the electric field of the radiation, which is
assumed uniform in the graphene plane in the geometry of normal
incidence.
The distribution function can be expanded in series in the electric field as follows
\begin{equation}
f(\bm{k},x,t) = f_0(\varepsilon_{\bm{k}}) + [f_1 (\bm{k},x)e^{- \mathrm i \omega t} + {\rm c.c.}]
+ f_2(\bm{k},x) + ... \:,
\end{equation}
where $f_0(\varepsilon_{\bm{k}})$ is the equilibrium distribution
function, $f_1 \propto |\bm{E}|$, and $f_2 \propto |\bm{E}|^2$.
The directed electric current along the structure edge is then
given by
\begin{equation}\label{J_edge_gen}
J_y = 4 e \int_{0}^{\infty} dx \sum_{\bm{k}} f_2(\bm{k},x) v_y \:.
\end{equation}
Here, the factor 4 accounts for the spin and valley degeneracy.

As a model, we consider the simplest form of collision integral,
\begin{equation}
Q \{ f(\bm{k},x,t) \} = - \frac{f(\bm{k},x,t) - f_0(\varepsilon_{\bm{k}})}{\tau},
\end{equation}
and assume the diffusive elastic scattering of carriers at the
edge, which implies that $f(\bm{k},0,t)$ is independent of the
direction of $\bm{k}$ for $v_x>0$ and $\int  f(\bm{k},0,t) v_x \,
d \varphi_{\bm{k}} = 0$.

To first order in the electric field, solution of
Eq.~(\ref{f_general}) with the above boundary conditions has the
form
\begin{multline}
f_1 (\bm{k},x) = - \frac{e \tau f'_0}{1-\rm{i}\omega\tau} \left[ \bm{E}_0\cdot\bm{v}  \right.\\
\left. - \left( \bm{E}_0\cdot\bm{v} + \frac{\pi}{4} E_{0,x} v \right)
\exp \left( - \frac{\rm{1}-i\omega\tau}{v_x \tau}x \right) \Theta (v_x) \right] \:,
\end{multline}
where $f'_0 = d f_0(\varepsilon)/d \varepsilon$ and $\Theta(v_x)$
is the step function equal to $1$ and $0$ for $v_x > 0$ and $v_x <
0$, respectively.

The equation for the second-order correction $f_2(\bm{k},x)$ to
the distribution function, which gives rise to $dc$ electric
current, assumes the form
\begin{equation}\label{equation_f2}
v_x  \frac{\partial f_2(\bm{k},x)}{\partial x}  + \frac{2e}{\hbar} {\rm Re}
\left[ \bm{E}_0^* \frac{\partial f_1(\bm{k},x)}{\partial \bm{k}} \right] = - \frac{f_2(\bm{k},x)}{\tau} \:,
\end{equation}
which yields
\begin{multline}\label{f2_dx}
\int_{0}^{\infty} f_2(\bm{k},x) dx = v_x \tau [f_2(\bm{k},0) - f_2(\bm{k},\infty)] \\
-  \frac{2e \tau}{\hbar} \int_{0}^{\infty} {\rm Re} \left[
\bm{E}_0^* \frac{f_1(\bm{k},x)}{d \bm{k}} \right] dx  \:.
\end{multline}
By using Eqs.~(\ref{J_edge_gen}) and~(\ref{f2_dx}) we derive for the edge electric current
\begin{equation}\label{J_y_2}
J_y = - 8 \frac{e^3 \tau^3}{\hbar} \sum_{\bm{k}} {\rm Re} \left\{ \frac{v_x v_y \bm{E}_0^*}{1-\rm{i}\omega \tau}
\frac{d [ (\bm{E}_0 \cdot \bm{v}) f'_0 ]}{d\bm{k}} \right.
\end{equation}
\[
\left. + \frac{v_y \bm{E}_0^*}{(1-{\rm i} \omega \tau)^2} \frac{d}{d\bm{k}} \left[
(\bm{E}_0 \cdot \bm{v} + \frac{\pi}{4} E_{0,x} v ) f'_0 \,v_x \right] \right\}  \Theta(v_x)  \:,
\]
%
where the above two contributions to the current stem from the
first and second terms on the right-hand side of
Eq.~(\ref{f2_dx}), respectively.

In graphene, the electron kinetic energy and velocity are given by
$\varepsilon_{\bm{k}} = \hbar v |\bm{k}|$ and $\bm{v} = v \,
\bm{k}/k$, respectively. Therefore, we obtain for the edge
photocurrent contribution from the electrons
\begin{multline}\label{J_y_3}
J_y = - \frac{e^3 \tau^3 v^2 f_0(0)}{2\pi \hbar^2  [1+(\omega\tau)^2]}
\left[ \left( 1 + \frac{7}{6} \, \frac{1-(\omega\tau)^2}{1+(\omega\tau)^2} \right) \right. \\
\left. \times (E_{0,x} E_{0,y}^* + E_{0,y} E_{0,x}^*) +
\frac{10}{3} \, \frac{\omega\tau}{1+(\omega\tau)^2} {\rm i}
[\bm{E}_0 \times \bm{E}_0^*]_z \right] \:.
\end{multline}
Similarly to the case of the photon drag effect the valence band holes contribution to the photocurrent can be obtained from  Eq.~(\ref{J_y_3}) by
the replacement $e \rightarrow -e$ and $f_0(0)  \rightarrow
f_0^{(\rm hole)}(0) = [1 - f_0(0)]$. Taking into account
both electron and hole currents we obtain that the total electric
current emerging at the sample edge is given by Eq.~(\ref{J_y_3}),
where $f_0(0)$ should be replaced by $[2f_0(0)-1]$.

%

The dependences of the edge photocurrent~(\ref{J_y_3}) on the
angles $\varphi$ and $\alpha$ are given by
\begin{equation}
\label{jy:simp:circ}
J_y(\varphi) \propto \frac{C_l}{2} \sin{4\varphi} + C_c \sin{2\varphi},
\end{equation}
\begin{equation}
\label{jy:simp:lin}
J_y(\alpha) \propto C_l \sin{2\alpha}.
\end{equation}
The microscopic theory for the edge photocurrents is in a good
agreement with experimental results. At normal incidence the
observed photoresponse is, according to Eq.~\eqref{jy:simp:circ},
a sum of the contributions proportional to function
$\sin{2\varphi}$ and $B \sin{(4\varphi)} + C \cos{(4\varphi)}
\propto \sin{(4\varphi+\xi)}$, see Fig.~\ref{fig3}(a). The
additional phase $\xi$ in experimental dependences results from
the arbitrary orientation of the sample edges. Moreover, the
dependence of the experimental data of the azimuthal angle
$\alpha$, Fig.~\ref{fig3}(b), demonstrates $\sin{2\alpha}$ and
$\cos{2\alpha}$ behavior in agreement with
Eq.~\eqref{jy:simp:lin}. Hence, helicity driven and linear
polarization dependent photocurrents in graphene are well
described at the normal incidence by our model of edge effects.

\section{Summary}

To summarize, our observations clearly demonstrate that the
irradiation of monolayer graphene flakes results in  directed
electric currents of different origins. In all our measurements a
substantial contribution to the photocurrent is driven by the
photon helicity. It can be separated into the contribution resulting
from normal incidence and the one from oblique incidence. While
the contribution from oblique incidence is related to the bulk
material and results from the transfer of the photon angular and
linear momentum to free carriers, the effect at normal incidence
is caused by the sample edges and vanishes in the bulk material.
Our theory describes the helicity driven as well as the
linear-polarization driven photocurrents in the classical limit
where the radiation frequency is smaller compared to the
characteristic energy of the carriers. The treatment of the
general case, where interband transitions should be taken into
account, is a future task.

\emph{Note added.} At the day of the submission of this manuscript, Entin et al. have published a pre-print\cite{Entin:gr} devoted to the theory of the linear photon drag effect in graphene. Unlike in Ref.~\onlinecite{Entin:gr}, where the direct interband absorption of linearly polarized light is considered, here we present results of mutual experimental and theoretical studies of the photon drag effect in graphene under indirect intrasubband optical transitions for both linear and circular polarizations, with emphasis on helicity-dependent photocurrents.

\acknowledgments We thank J.~Fabian, V.V.~Bel'kov, J. Kamann, and
V. Lechner for fruitful discussions. The  support from the DFG,
the Linkage Grant of IB of BMBF at DLR, RFBR, President grant for
young scientists and the ``Dynasty'' Foundation -- ICFPM is
gratefully acknowledged.


\begin{thebibliography}{99}
\bibitem{footnote1} In spite of the fact that care was taken to align the radiation propagation
direction most closely parallel to the line connecting the contacts the slight deviations are
possible since the invisible radiation and the
small sample size prevent a full control of the geometry.
\bibitem{footnote2} We note that
ripples and other imperfections generally can also reduce the symmetry and yield a photocurrent.
In our samples, however, AFM
images do not demonstrate the presence of ripples, and, to our opinion,
it is likely that the effect comes from edges.
\bibitem{footnote3}  We note that the microscopic mechanism of the edge current
is  similar to the surface photogalvanic effect observed in
bulk materials~\protect\cite{gurevich,Magarill,Alperovich}.
\end{thebibliography}
\end{document}